# Tone generation in an open-end organ pipe:
# How a resonating sphere of air stops the pipe

Bernhardt H. Edskes[1+], David T. Heider[2+], Johan L. van Leeuwen[3], Bernhard U. Seeber[4], and J. Leo van Hemmen[2*]

**Abstract** According to the classical Helmholtz picture, a flue organ pipe while generating its eigentone has two anti-nodes at the two open ends of a cylinder, the anti-nodes being taken as boundary condition for the corresponding sound. Since 1860 it is also known that according to the classical picture the pipe actually sounds lower, which is to say the pipe "sounds longer" than it is, for long a physical puzzle. As for the pipe's end, we have resolved this acoustic enigma by detailing the physics of the airflow at the pipe's open end and showing that the boundary configuration is actually the pipe's acoustically resonating vortical sphere (PARVS). The PARVS geometry entails a sound-radiating hemisphere based on the pipe's open end and enclosing a vortex ring. In this way we obtain not only a physical explanation of sound radiation from the organ-pipe's open end, in particular, of its puzzling dependence upon the pipe's radius, but also an appreciation of it as realization of the sound of the flute, mankind's oldest musical instrument.

## I. INTRODUCTION

With a respectable age of at least 35,000 years, the bone flute[1-3] is one of humanity's oldest, presumably even the oldest, musical instrument. Reduced to the bare essentials[1], it is a thin tube with a sound-generating mouthpiece to start with, a few holes in between that allow for producing different tones, and an open end. Here we analyze the air configuration at the open end during flute playing, focus on the organ pipe as an also respectably aged reduction of the problem to its mere essence, and present an end configuration that is rather different from what one, since long[2-14], had accepted.

Though an order of magnitude in millennia younger, the pipe organ[4-6] as composite of many pipes and, hence, tones, is also of a respectable age (3rd century BCE). Metallic open-end flue pipes are the majority; cf. Fig. 1a-c. Inspired by clear physical imagination (see Fig. 1d), Helmholtz[7] was the first to mathematically derive a relation between a pipe's fundamental frequency $f_0$ and the pipe length $L$ in that $f_0 = 0.5\, c/L$ with $c$ as sound velocity in air. The underlying physics of having two anti-nodes separated by $L$ is appealing but turned out incomplete. It was found[2-13] that one needs to replace $L$ by $L + (\Delta_e + \Delta_m)$ where $\Delta_e = 0.6\, a$ is the end correction, with $a$ as the pipe's radius and $f_0$ small enough[13], and $\Delta_m$ is the labial correction for the mouth[14]; most of $\Delta_m$'s analytical calculation is still inaccessible. Scientific focus has been on $\Delta_e$ ever since[2-13]. Here, the

---

[1]Edskes Orgelbau, Wohlen, Switzerland.

[2]Physik Department T35, Technical University of Munich, Garching bei München, Germany.

[+] These authors contributed equally to the PARVS discovery.

[3]Experimental Zoology Group, Department of Animal Sciences, Wageningen University, Wageningen, The Netherlands.

[4]Audio Information Processing, Technical University of Munich, München, Germany

[*] Corresponding author. Email: lvh@tum.de

puzzling presence of *a* in the end correction $\Delta_e = 0.6\,a$ is explained and interpreted physically: Only during tone generation does an end correction arise from the pipe's acoustically resonating vortical sphere (PARVS; see Figs. 1f, 2 and 3) with radius *a* at the pipe's end in conjunction with a robust vortex ring. Experimental verification has been performed at a usual blowing pressure of 7 cm water and visualization has been realized classically through cigarette smoke[15].

## II. ENSUING DEVELOPMENTS AND PRESENT EXPERIMENTS

Organ pipes are blown at a very low pressure *p* of 5-10 cm water (as compared to nearly 10 m water for atmospheric pressure) allowing the upper lip (Fig. 1a&b) to produce[2-6] a broad sound spectrum that excites the pipe's eigenfrequencies; cf. Fig. 4. Computing an organ pipe's dominating fundamental frequency $f_0$ for a given pipe length *L* has tantalized science since the organ's more frequent appearance[6] in the Middle Ages. Only as late as 1860 did Helmholtz[7] mathematically derive a comprehensive theory including $f_0$ for an open-end flue pipe from first principles; see Fig. 1d. Newton and presumably also Huygens were his predecessors[6] but Helmholtz' derivation was the first to be complete. The two ends, i.e., below the mouth's upper rim just above the flue and at the top, are open so that during resonance an anti-node ought to be here. Sound being a longitudinal wave (for its geometry, see Fig. 1d and below), $\lambda/2 = L$ for the key tone and thus we obtain the fundamental frequency $f_0 = 0.5\,c/L$, based on simple geometry that can be translated into a (Neumann) boundary condition with vanishing vertical derivative of the air's velocity. Rayleigh[9,10] replaced it by a flat massless, fictitious, piston that was used to compute the ensuing sound radiation.

To understand the physics underlying Helmholtz's creative imagination, we must make a small detour. The flue blows the air against the rim of the upper lip, thereby generates a broadband sound spectrum, and in this way excites the pipe's eigenmodes[2-6]; cf. Fig. 4. Since the fundamental tone's wavelength $\lambda$ equals the velocity of sound *c* divided by the fundamental frequency $f_0$ and $\lambda/2 = L$, for the fundamental frequency we end up with $\lambda = 2L = c/f_0$ so that $f_0 = 0.5\,c/L$. Already as early as 1860 did Cavaillé-Coll[8] publish an experimental paper showing that the actual fundamental frequency is lower than $0.5\,c/L$ or, equivalently, that the so-called effective pipe length is longer than *L*. This effective length $L_{eff}$ to compute $f_0$ through $\lambda = 2\,L_{eff} = c/f_0$ has therefore been taken to be $L_{eff} = L + \Delta_m + \Delta_e$ with positive phenomenological correction factors $\Delta_e$ and $\Delta_m$ to take care of end and mouth corrections, whose physical interpretation was nonexistent. Both the complicated geometry of and the turbulence at the rim of upper lip still preclude[3,14] any simple physical explanation of $\Delta_m$. We therefore focus on $\Delta_e$. Where, then, does the end correction $\Delta_e$ come from? As Figs. 2 and 3 show, the pipe's acoustically resonating vortical sphere (PARVS) physically explains the end correction $\Delta_e$, which is a radical break with history[2-13]. To see this, we return to history for a moment.

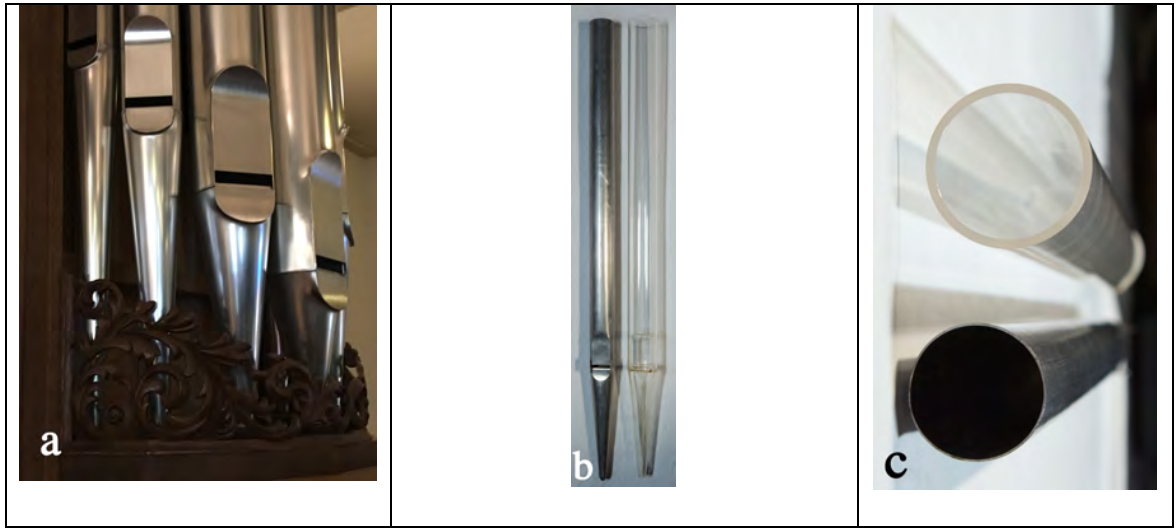

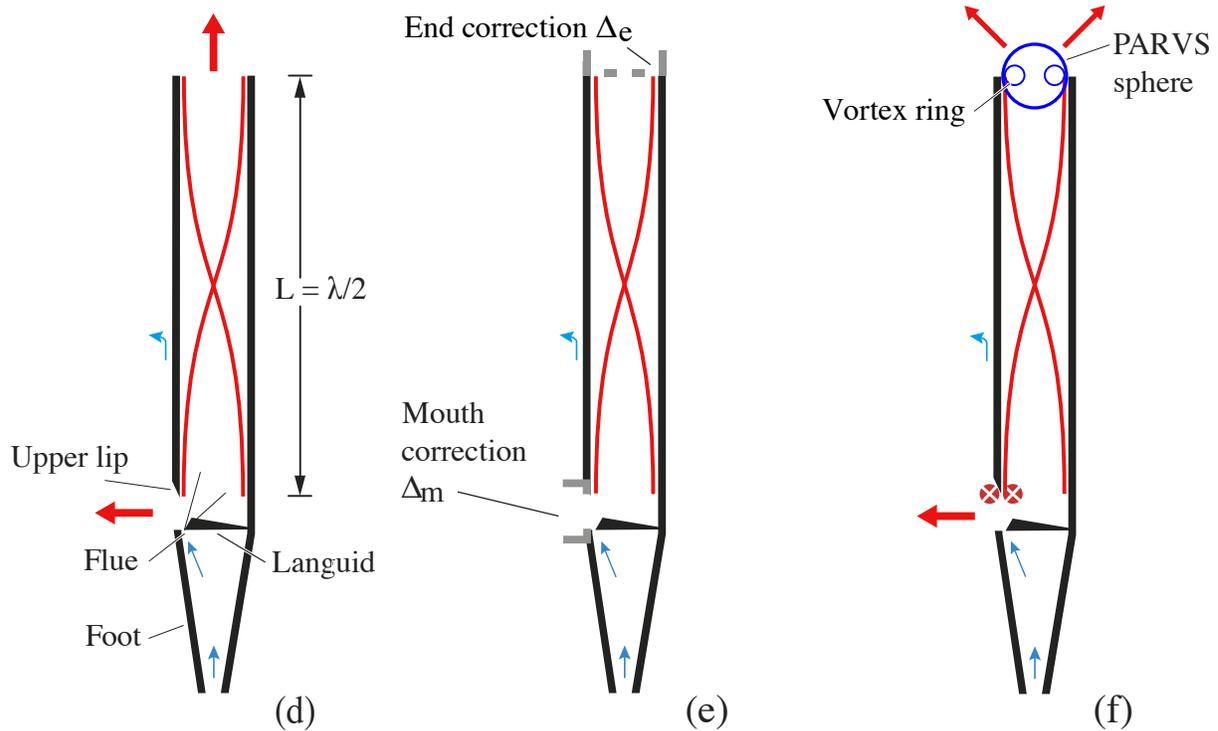

**Fig. 1. Organ pipes & physics underlying an open-end organ pipe generating its fundamental frequency. a** Circular flue (or labial) pipes in an organ, the majority. The middle pipes show the mouth, the rectangular opening (black) between upper and lower lip. A tone is generated once an air stream stemming from the flue underneath it hits the rim at the bottom of the upper lip; cf. (**d**). **b** Pipes as used here in experiment; left a regular tin pipe (70% Sn, 30% Pb), right a plexiglass one for visually verifying phenomena in the former. **c** Same pipes as in (**b**) but seen from above. Dimensions: Length $L$ = 56.2 cm (cf. Fig. 2a), inner diameter $D$ = 40 and 44 mm, and wall thicknes $d$ = 0.8 and 3.0 mm, respectively. **d** Sound-generation mechanism according to Helmholtz[7], schematic. For an open-end organ pipe, both the mouth and the end are open so that two anti-nodes exist there and, as the figure shows, $L = \lambda/2$. The pipe's fundamental tone has a single node in between. Sound is radiated at the top and through the mouth at the bottom (boldface red arrows).

**e** Rayleigh's calculation[10,11] of the end-correction through a fictitious massless piston (grey) is consistent with the Helmholtz picture[7] and indicates the pipe effectively becomes longer by $\Delta_e$. **f** In reality, the PARVS (blue sphere) stops the pipe. The pipe indeed becomes longer, by about $2a/3$ where $a = D/2$ is the pipe's radius (see text). Instead of Rayleigh's[10,11] massless, fictitious, flat piston in (**e**), the PARVS radiates sound as a hemisphere, indicated by the two red arrows. As in the flute, the upper lip is responsible for tone generation, symbolized by two crossed rolls (red).

Rayleigh[9] took up Helmholtz's lead[7] in 1871, assumed a fictitious, flat, massless, but vibrating piston as boundary condition at the pipe's end and showed it would emanate sound of a wavelength with $L/2$ replaced by $(L + \Delta_e)/2$ with $\Delta_e = 0.3\,a$. In 1877 Rayleigh reported[11] $\Delta_e = 0.6\,a$, and so did Bossanquet[12]. The correction problem becomes even more poignant in that a piston respects Helmholtz' imagination of Fig. 2a but the presence of the pipe's radius $a$ does not. And, by exception, Rayleigh's imagination did not quite agree with experiment. As shown in Figs. 2 and 3, a sphere and not Rayleigh's fictitious piston stops the pipe. The sphere's radius is $a$ while the sound-radiating hemisphere on top of the pipe (Fig. 1f) makes the pipe effectively longer so that $a$ naturally comes in to explain the length correction $\Delta_e$. The geometry of Fig. 1f directly reveals $\Delta_e < a$ and we will show $\Delta_e \approx 2a/3$.

While tuning a circular, gold-plated, organ pipe with well-lit rim, Bernhardt Edskes discovered the inner vortex of Fig. 1e through a small gold particle that was rotating "in the air" near the rim, while keeping its position. As we now see in Fig. 2, it was a vortex ring carrying the particle. Doing the underlying math – see Methods – it was found that in the limit $p \to 0$ (p ≈ 7 cm H$_2$O; as shown below, p$_{end}$ at the top is even far less) a sphere in conjunction with a vortex-ring structure appears; cf. Figs. 2 & 3. One may therefore call this configuration the pipe's acoustically resonating vortical sphere (PARVS). As confirmed by experiment, it only occurs during tone generation; see the videos.

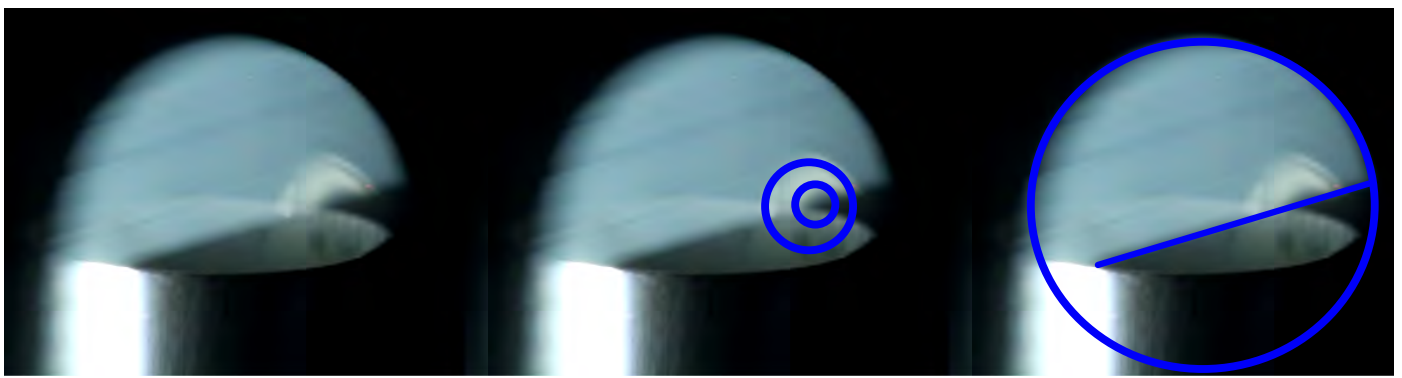

**Fig. 2. Pipe's acoustically resonating vortical sphere** (PARVS, blue on the far right) as well as the inner vortex-ring structure (blue, in the middle) at the end of a tin organ pipe with fundamental tone of 257 Hz (Figs. 1 & 4) and length $L$ = 56 cm have been made visible by means of a thin vertical "light sheet" due to a diascope at a distance of 2 m and slicing the cigarette smoke[15] that stems from the pipe's foot. In the middle image the inner vortex ring has been accentuated. The PARVS makes, so to speak, the pipe longer and the correction factor $\Delta_e$ needs to include the PARVS's radius $a$, which is also the radius of the pipe. Hence an end-correction of $\Delta_e \approx 0.6\,a$ results, which is to be added to the pipe length $L$ as the PARVS effectively elongates the pipe's resonance body; cf. Fig. 1f.

How can we understand $\Delta_e = 2a/3$ intuitively? The PARVS's sound-radiating hemisphere (Figs. 1f and 2) comes atop the cylindrical organ pipe. Replace the hemisphere by a cylinder with an effective height $\Delta_e$ so that the two volumes are equal, $(2\pi/3) a^3 = (\pi a^2) \Delta_e$. Hence $\Delta_e = 2a/3$ is to be added to $L$ in Fig. 1d. The prefactor 2/3 is in agreement with experiment[16] and up to 10% with the theoretical 0.61 (for $ka \ll 1$ with $k = 2\pi/\lambda$) due to Levine and Schwinger[13], who performed a Wiener-Hopf *tour-de-force* to obtain their result.

Because at the flue (Fig. 1d) there is a slight overpressure $p$ (of the order of 7 cm $H_2O$) blowing the pipe, one might argue that vortices are always present[17] at the pipe's end. The labium's upper lip (Fig. 1) has therefore been covered (by chewing gum) so that the pipe was blown but the wedge was gone, no tone was produced, and no PARVS appeared. Nevertheless, the resonance locking of Helmholtz[7] for the sounding pipe as shown in Fig. 1d in conjunction with the flow physics of e.g. Krutzsch[17] intuitively suggests their combination as a physically consistent picture; hence the name of PARVS. For $p > 0$, the vortex ring is hovering slightly above the pipe's rim (Fig. 3a). And in contrast to classical flow physics[17], the vortex ring is stable. That is, it does not move upwards and is only present during tone-generating resonance.

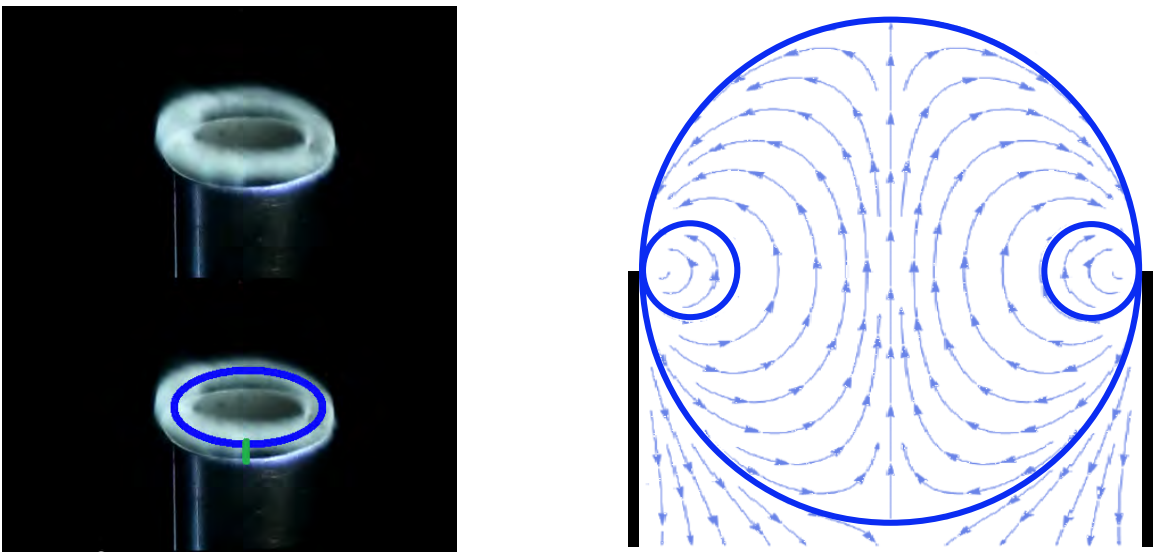

**Fig. 3. PARVS's vortex ring and sphere.** At the end of the same tin pipe as in Fig. 2 one sees a vortex ring (left), accentuated by a blue ring. Right: A mathematically calculated snapshot (see Methods) shows a vortex ring merging into a sphere at the PARVS's boundary. Both are visible due to a high enough cigarette-smoke concentration[15] at the pipe's end. Bottom left: As indicated by the (because of visualization too long) vertical green line segment, the vortex structure is hovering slightly above the pipe's open end because of the small overpressure due to blowing, which at the pipe's end amounts to $p_{end} < 0.1$ mm ≈ 1 Pa. In comparison, on the right: mathematical model (2-dimensional cut, instantaneous) under ideal conditions, i.e., in the limit of vanishing blowing pressure ($p \to 0$). One sees both the blue PARVS sphere and the – much smaller – vortex ring it contains.

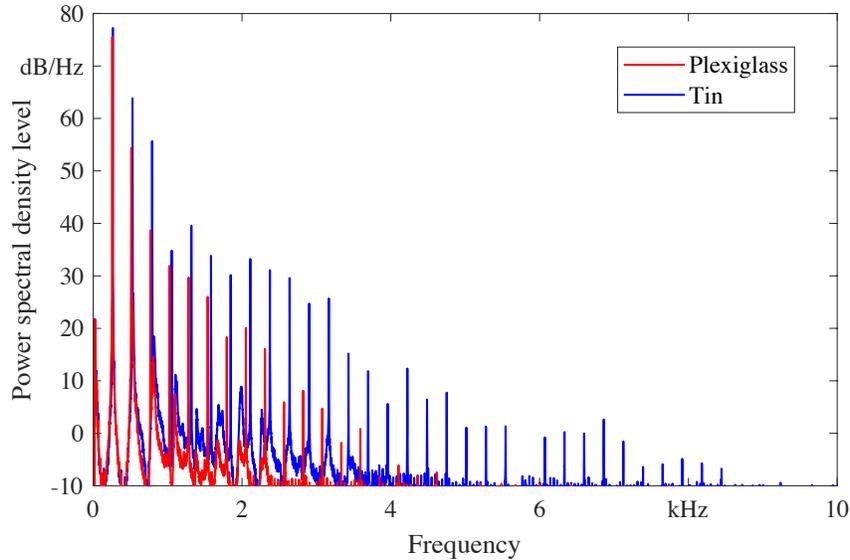

**Fig. 4. Spectral decomposition of organ pipe' acoustic output.** For the tin and plexiglass pipes (Figs. 1b&c) as used here, the plot shows the fundamental tone (experimental $f_0$ of 264 Hz and 257 Hz, respectively, to be compared with $f_0$ = 0.5 $c/L$ ≈ 307 Hz for c ≈ 345 m/s, and 274 Hz according to Cavaillé-Coll's estimate[8] of $L_{eff}$ ≈ $L + 10a/3$) as well as higher harmonics, measured in an anechoic room; see Methods.

## III. RESULTS AND DISCUSSION

Let us step back for an overview before turning to conclusion and outlook. Extreme care was taken to equip the pipe with its *natural* frequency. Air enters an organ pipe – and any reedless wind instrument – through its flowing across a usually rectangular slit at one end, here the flue at the bottom of the pipe, as shown in Figs. 1a and 1d, and hits a sharp edge, here the labial upper lip. The air pressure is macroscopically rather low, 5-10 cm water, but compared to the emitted sound oscillations extremely high. As the air hits the sharp edge of the upper lip (Fig. 1d), vortex-like structures arise, which have already been found and published as early as 1938 by Burniston Brown[15], who used cigarette smoke to visualize them. They have been studied ever since; see e.g. Rienstra and Hirschberg[18]. These vortex-like phenomena are not pure vortices but vortex-*like* structures whose intricacy hints at their inherently chaotic character. They all arise at the upper lip of the organ pipe or flute[1-3] or whatever reedless wind instrument[2-5,18], produce a broad spectrum so as to excite the instrument's natural vibration with its specific fundamental frequency and overtones, and are as far away as possible from what happens at the instrument's pipe ending, object of the present study. In fact, the PARVS is of a completely different nature, it is deterministic and does not exhibit any chaos whatsoever.

Nevertheless a first reaction might be that Laser-Doppler Anemometry (LDA) or Particle-Image Velocimetry (PIV) ought to be done so as to confirm the present results exhibiting the PARVS. Not only is cigarette smoke a classic experimental visualization[15], generally accepted and also used here, but neither LDA nor PIV can, in fact, do so. LDA cannot show the PARVS as smoke particles move too slowly to allow its functioning whereas for PIV they are in practice far too small. Nice PIV experiments[19] have been done at the other end of an organ pipe, viz., the flue, at a regular, three orders of magnitude higher flue pressure of 5-10 cm water, and only exhibit the flue flow mentioned above. Open-end organ-pipe experiments[20] using PIV have been performed at blowing pressures exceeding by far the ones that govern the regular performance of an organ pipe, viz., 5-10 cm water pressure; as little as < 20% higher pressure already leads to overblowing and the first overtone, a few % more to the second overtone, etc. Much higher pressures $p$ forbid[2-6] the pipe's normal functioning but are needed for modern PIV. The resulting acoustic output is in the range of[20] 150 dB SPL (Sound Pressure Level) and, hence, of nonlinear acoustics, in contrast to the 60-100 dB SPL for normally functioning organ pipes, whose asymptotic behavior is fully covered by linear acoustics. For instance, Fig. 2's pipe had an output of 75 dB SPL at its top; see Methods.

Given the fact that both the oscillation-resonance pressure of sound and the overpressure $p_{end}$ at the pipe's end are of the same order of magnitude (< 1 Pa ≈ 0.1 mm $H_2O$) and three orders of magnitude less than at the other end of the pipe, at the flue, we have taken recourse to a classic cigarette-smoke technique[15], with extremely light smoke particles of size < 0.3 $\mu$m (mean value[21,22]) so as to experimentally confirm and visualize our findings: Instead of a fictitious vibrating piston, the end correction $\Delta_e$ is due to a vibrating sphere consisting of resonating air and stopping the pipe, viz., the pipe's acoustically resonating vortical sphere (PARVS) in conjunction with a stable vortex-ring inside the sphere; see Figs. 2 & 3, as well as Fig. 4 for the ensuing spectral decomposition. The radii of sphere and pipe are identical. We therefore obtain an end correction depending on their common value $a$ and in this way resolve a long-standing physical enigma.

From a physical perspective, we now have both a clear geometrical picture of what generates the pipe's end correction and the mathematical setup for the corresponding boundary condition, viz., the very same PARVS, to compute the emanating sound radiation. From first principles, neither the pipe's fundamental frequency nor the overtones have ever been calculated analytically. Instead, boundary conditions as depicted by Fig. 1d were used as an *a priori* ansatz. Helmholtz's original choice[7] looks highly plausible as the open end "evidently" stands for an anti-node. Rayleigh[9-11] then used this physical imagination to replace it by a fictitious massless piston to *compute* the emanating sound radiation. We now know that these ideas, fruitful as they were, need at least to be complemented by the PARVS. Currently its primary justification is experimental visualization through a classic technique.

From a conceptual perspective, the PARVS' origin exemplifies the deep and close relationship between the geometry of wind instruments – here, the organ pipe – and their acoustic properties. It may well be worthwhile to assess whether and, if so, to what extent the notion of PARVS discovered in the context of organ pipes can be extended so as to explain potentially similar resonance properties of other open-ended wind instruments, reedless or not.

## IV. Methods

### A. PARVS as a novel boundary condition?

Figure 1 visualizes in 1d the Helmholtz model, in 1e known corrections $\Delta_m$ and $\Delta_e$, and in 1f the geometrical explanation induced by the PARVS that is presented for the end-correction $\Delta_e$. We now focus on a straightforward mathematical description of what a sphere in conjunction with a system of vortex rings, which together constitute the PARVS and stop the pipe, looks like.

As Figs. 3 and 4 show, a discussion of PARVS's existence and stability is moot. Motivated by PARVS's discovery as a consequence of a stable vortex-ring system at the pipe end, we simply present an ansatz – as does any other mathematical expression describing a specific vortex configuration, such as those named after Rankine and Hill[23]. Since the ring is rotationally invariant we restrict ourselves to a two-dimensional problem in the $(x, z)$-plane with the pipe end at $z = 0$ and take two vortex dipoles of strength $\Gamma$, symmetrically positioned w.r.t. the (vertical) z-axis in the $(x, z)$-plane. That is, our ansatz for the air-velocity vector $\mathbf{v}(x, z) = (v_x, v_z)$ in the $(x, z)$-plane is given by

$$\mathbf{v}(x,z) = (v_x, v_z) = \sum_{i,j=0}^{1} \frac{\Gamma}{2\pi} \frac{(-1)^{i+j} r_{\text{wall}}}{x_{ij}^2 + z^2} (x_{ij}, z), \quad (1)$$

$$\text{with} \quad x_{ij} = x - (-1)^i \left[ a + (-1)^j r_{\text{wall}} \right]. \quad (2)$$

Experimentally, the outside vortex ring has turned out to be unstable and, hence, absent. The pipe's wall thickness is taken to be $d = 2r_{\text{wall}}$ so that the pipe walls are between $r = a \pm r_{\text{wall}}$ while the inner diameter $D = 2(a - r_{\text{wall}})$; cf. Fig. 1. The prefactor $r_{\text{wall}}$ in the numerator of Eq. (1) effectuates the outcome of (1) becoming well-defined in the limit $r_{\text{wall}} \to 0$, with $v_x = 0$ at $r = (a - r_{\text{wall}})$ and $z < 0$. Consequently, (1) gives rise to a description of the PARVS as a vibrating quadrupole[23,24] that has been used to compute the emergent sound radiation and cover the experimental results. Details are to be given elsewhere because their mathematical intricacy goes beyond the scope of the present paper. From a more elevated point of view, one can say that, in the limit $r_{\text{wall}} \to 0$, the PARVS stops the open end of a cylindrical organ pipe as a consequence of the topological irregularity at the pipe's rim, viz., the metal-air discontinuity. The richness of phenomena for the opposite limit $r_{\text{wall}} \gg a$ remains to be uncovered, though.

In Fig. 3b everything plotted is effectively dimensionless with $a$ taken to be $a = 1$. Moreover, $r_{\text{wall}}/a = 1/17 \ll 1$ so as to represent a thin pipe wall of thickness $2r_{\text{wall}}$. See also Fig. 1 with $a = 20$ mm, $d = 2r_{\text{wall}} = 0.8$ mm and $r_{\text{wall}}/a = 1/50$ so that there is nothing against identifying the main text's $a$ and $a - r_{\text{wall}}$ here, the difference being totally indiscernible in experiment. Figure 3b shows the resulting PARVS at a certain moment, which is to stop the pipe's resonating eigenmodes. Accordingly, only configurations inside and on top of the pipe are to be taken into account.

## B. Experimental setup: Pipes, optics, and audioacoustics

**Organ pipes**. The organ pipes have been specially built by Edskes Orgelbau, Wohlen, Switzerland. Their dimensions and compositions as alloy have already been specified in the main text.

**Optics**. The organ pipes shown in Fig. 1 were installed perpendicularly to the laboratory floor and connected to an air-bag system supplying a practically constant pressure of $p = 7$ cm water. In Fig. 5, the tube system for the airflow is indicated in green. The lab was darkened and a diascope illuminated a black slide with only a single slit of width $< 0.1$ mm with orientation vertical or parallel to the horizontal plane through the pipe end's rim so as to "slice" the smoke at the pipe top; see, for instance, Fig. 3. A suction pump was connected to both 1–2 cigarettes and the air supply system to inject tobacco smoke into the pipe through the foot hole; cf. Fig. 5. A Canon 6D full-format digital camera with 24-70 mm/f2.8 lens was used to film the evolution of the PARVS and its vortex structure.

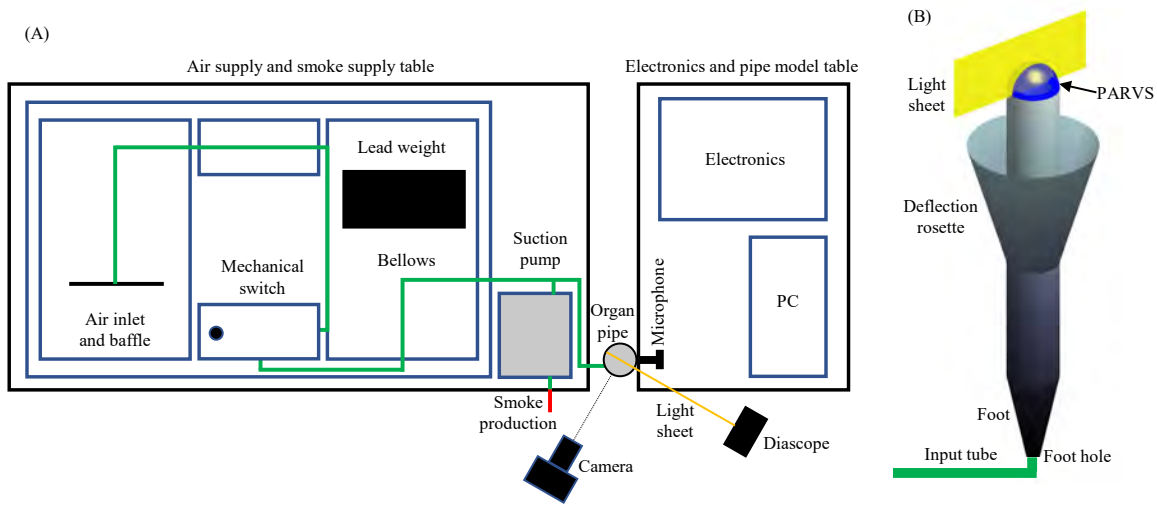

**Fig. 5. Experimental setup, schematic.** (A) Scheme of the diverse apparatus. (B) Rosette to guarantee that open-end phenomena at the top, particularly, the PARVS, are not perturbed by air emission from the labium. The weight that one can shift up and down (left) allows a continuous regulation of the pipe's blowing pressure so as to reach its required fine-tuning for resonance.

**Audioacoustics.** The optical setup was combined with two microphones, one placed at a distance of 30 cm measured w.r.t. the center of the open end of the pipe along a line making an angle of 45° with the vertical so as to record the radiated sound while another microphone was positioned 30 cm directly in front of the labium. Free-field pressure microphones NTI M2230, class 1, were used. Recording was via RME Micstacy pre-amplifier with 30 dB gain and RME Babyface Pro sound card into a laptop with Audacity software at a sampling frequency of $f_s = 96$ kHz and 24-bit resolution.

**Videos.** Experimental verification of the PARVS is nontrivial. For instance, where it appears, the smoke particles are to get caught by the resonance pattern as already shown by Figs. 3 and 4. It should, however, be constantly borne in mind that the PARVS amplitudes at the pipe's end are very small, < 100 nm. Even if someone walks through the lab very slowly, which is necessary every now and then, a huge momentum of air is induced that effortlessly drags the (macroscopic) smoke particles à la Stokes[23, 25] and dominates the phenomena we are after. Accordingly, the return of the surrounding air back to rest takes extremely long. The four videos exhibit both the PARVS sphere and the vortex-ring, or both, as displayed by Figs. 3 & 4, or none once the tone is absent but the flue stream is still blowing. The schematic of the experimental setup is shown in Fig. 5.

## Supplementary Information

Supplementary information, i.e., videos: The online version contains supplementary material available at…

## Acknowledgments

The authors thank Sepp Kressierer (TUM) for providing a suitable suction pump, Bruce A. Young (A.T. Still University, Kirksville, MO) for helpful advice and Sjoerd Rienstra (TU Eindhoven) for a critical reading of the manuscript.

## Author contributions

B.H.E. discovered a gold particle's floating in air at the pipe's top during tone generation, J.L.v.H. realized the vortex ring underlying it, conceived and supervised the whole project, and wrote the manuscript. D.T.H. developed the mathematical theory and in so doing discovered the PARVS (= vortex ring **+** sphere). J.L.v.L. conceived the light slicer stemming from a diascope, and B.U.S. performed the acoustical measurements. B.H.E. is responsible for the manufacture of the organ pipes. All authors participated in the experiments, discussed the results, and commented the manuscript.

## Conflicts of interests

The authors have no conflicts to disclose.